\documentclass[10pt,twocolumn,superscriptaddress,preprintnumbers,showpacs]{revtex4}
\usepackage{epsf}
\usepackage{amsmath}
\usepackage{amssymb}
\usepackage{epsfig}
\usepackage{latexsym}
\usepackage{amsfonts}
\usepackage{varioref}
\usepackage{ifthen}
\usepackage{graphicx}
\usepackage{amssymb,amsmath,amsthm}%
\providecommand{\U}[1]{\protect\rule{.1in}{.1in}}
\begin{document}
\parindent 0mm 
\parindent 0mm 
\setlength{\parskip}{\baselineskip}
\pagenumbering{arabic} 
\setcounter{page}{1}
\mbox{ }
\preprint{UCT-TP-289/12}
\title{Chiral symmetry restoration and deconfinement in QCD at finite temperature}
\vspace{.1cm}
\author{C. A. Dominguez}
\affiliation{Centre for Theoretical \& Mathematical Physics, University of Cape Town,
Rondebosch 7700, South Africa}

\author{M. Loewe}
\affiliation{Centre for Theoretical \& Mathematical Physics, University of Cape Town,
Rondebosch 7700, South Africa}
\affiliation{Facultad de F\'{i}sica, Pontificia Universidad Cat\'{o}lica de Chile, Casilla 306, Santiago 22, Chile}

\author{Y. Zhang}
\affiliation{Centre for Theoretical \& Mathematical Physics, University of Cape Town,
Rondebosch 7700, South Africa}
\date{\today}
\begin{abstract}
\noindent 
The light-quark correlator in the axial-vector channel is used, in conjunction with finite energy QCD sum rules at finite temperature, in order to (a) establish a relation between chiral-symmetry restoration and deconfinement, and (b) determine the temperature behavior of the $a_1(1260)$ width and coupling. Results indicate that deconfinement takes place at a slightly lower temperature than chiral-symmetry restoration, although this difference is not significant given the accuracy of the method. The behaviour of the $a_1(1260)$ parameters is consistent with quark-gluon deconfinement, as the width grows and the coupling decreases with increasing temperature.\\
\end{abstract}
\maketitle
\noindent

\section{Introduction}

The extension of the QCD sum rule method \cite{REVIEW} to finite temperature was first proposed in \cite{BS}, and applied to the light-quark vector meson system, i.e. the $\rho$-meson channel. Additional theoretical support for the validity of such an extension was  provided later in \cite{OPET}.
A quark-gluon deconfinement parameter was also introduced in \cite{BS} in the form of the squared energy threshold, $s_0(T)$, for the onset of perturbative QCD (PQCD) in hadronic spectral functions. Around this energy, and at zero temperature, the resonance peaks in the spectrum are either no longer present or become very broad. The smooth hadronic spectral function thus approaches the PQCD regime.
With increasing temperatures approaching the critical temperature for deconfinement, $T_c$, one would expect hadrons to disappear from the spectral function which should then be described entirely by PQCD. The analysis of \cite{BS} indeed showed $s_0(T)$ to be a monotonically decreasing function of the temperature, together with the coupling of the $\rho$-meson to the vector current. Since this analysis was performed in the zero-width approximation, an important dynamical feature also signaling deconfinement was overlooked, i.e. resonance broadening as proposed in \cite{DL1}-\cite{DL2}, and subsequently confirmed in  other applications \cite{GAMMA3}.\\
A link between deconfinement and chiral-symmetry restoration using QCD sum rules in the axial-vector channel was first established in \cite{DL1}, improved in \cite{GATTO1}, and recently updated and extended to finite density in \cite{DL3}.
These analyses indicate that the temperature at which $s_0(T)$ vanishes is very close to that at which the quark condensate, or alternatively the pion decay constant $f_\pi(T)$ vanishes. Within the accuracy of the method these results imply that both phase transitions take place at roughly the same temperature.\\
The analyses of \cite{DL1}, \cite{GATTO1}-\cite{DL3} made use of the finite energy QCD sum rule (FESR) of the lowest dimension ($d=2$) in the axial-vector channel, assuming full saturation of the hadronic spectral function by the pion pole. This assumption would not be entirely justified if one were to consider the subsequent two FESR of dimension $d=4$ and $d=6$ in order to extract more information from the sum rules. In fact, already at $T=0$ one finds that the values of the condensates of dimension $d=4$ and $d=6$ that follow from the second and third FESR are barely consistent with results obtained using experimental data \cite{DS1}. This strongly suggests additional hadronic contributions, and in fact the data in this channel include not only the pion pole but also the $a_1(1260)$ resonance. A straightforward theoretical calculation confirms  this to be the case.\\

In this paper we reconsider the light-quark axial-vector channel using the first three FESR, together with an improved hadronic spectral function involving the pion pole as well as the $a_1(1260)$ resonance. This allows for a more realistic conclusion on the relation between chiral symmetry restoration and deconfinement. At the same time, it provides additional and valuable information on the temperature behaviour of the $a_1(1260)$ coupling and hadronic width. The results indicate that $s_0(T)$ vanishes at a critical temperature some 10\% below that for chiral-symmetry restoration. Within the accuracy of the method this difference is not significant. The $a_1(1260)$ coupling initially increases with increasing $T$ up to $T/T_c \simeq 0.7$, and then decreases sharply up to $T_c$. The hadronic width of the $a_1(1260)$ remains constant up to $T/T_c \simeq 0.6$, increasing sharply thereafter. This behaviour of the coupling and the width are fully consistent with a quark-gluon deconfinement scenario.
\\

\section{Finite Energy QCD Sum Rules at ${\bf{T=0}}$}
\noindent
We consider the correlator of light-quark axial-vector currents
\begin{eqnarray}
\Pi_{\mu\nu} (q^{2})   &=& i \, \int\; d^{4} \, x \; e^{i q x} \; 
<0|T( A_{\mu}(x) \;, \; A_{\nu}^{\dagger}(0))|0> \nonumber \\ [.3cm]
&=& -g_{\mu\nu}\, \Pi_1(q^2) + q_\mu q_\nu\, \Pi_0(q^2)  \; ,
\end{eqnarray}
where $A_\mu(x) = : \bar{d}(x) \gamma_\mu \, \gamma_5 u(x):$ is the (charged) axial-vector current, and $q_\mu = (\omega, \vec{q})$ is the four-momentum carried by the current. The functions $\Pi_{0,1}(q^2)$ are free of kinematical singularities, an important property needed in writing dispersion relations and sum rules. Concentrating on e.g. $\Pi_0(q^2)$ and invoking the Operator Product Expansion (OPE) of current correlators at short distances beyond perturbation theory, one of the two pillars of the QCD sum rule method, one has
\begin{equation}
4 \pi^2\,\Pi_0(q^2)|_{\mbox{\scriptsize{QCD}}} = C_0 \, \hat{I} + \sum_{N=1} \frac{C_{2N} (q^2,\mu^2)}{Q^{2N}} \langle \hat{\mathcal{O}}_{2N} (\mu^2) \rangle \;, \label{OPE}
\end{equation}
where $Q^2 \equiv - q^2$, $\langle \hat{\mathcal{O}}_{2N} (\mu^2) \rangle \equiv \langle0| \hat{\mathcal{O}}_{2N} (\mu^2)|0 \rangle$, $\mu^2$ is a renormalization scale, the Wilson coefficients $C_N$ depend on the Lorentz indexes and quantum numbers of the currents, and on the local gauge invariant operators ${\hat{\mathcal{O}}}_N$ built from the quark and gluon fields in the QCD Lagrangian. These operators are ordered by increasing dimensionality and the Wilson coefficients are calculable in PQCD. The unit operator above has dimension $d=0$ and $C_0 \hat{I}$ stands for the purely perturbative contribution. At $T=0$ the dimension $d=2$ term in the OPE cannot be constructed from gauge invariant operators built from the quark and gluon fields of QCD. In addition, there is no evidence from such a term from analyses  using the experimentally measured axial-vector spectral function \cite{DS1}. The dimension $d=4$ term, a renormalization group invariant quantity, is given by
\begin{equation}
C_4 \langle \hat{\mathcal{O}}_{4}  \rangle = 
\frac{\pi}{6} \langle \alpha_s G^2\rangle + 2 \pi^2 (m_u + m_d) \langle\bar{q} q \rangle , \label{C4}
\end{equation}
where the second term is negligible in comparison with the gluon condensate, and thus it will be ignored in the sequel.
The leading power correction of dimension $d=6$ is the four-quark condensate, which in the vacuum saturation approximation \cite{REVIEW} becomes
\begin{equation}
C_6 \langle \hat{\mathcal{O}}_{6}  \rangle = \frac{704}{81} \,\pi^3 \, \alpha_s \,|\langle \bar{q} q \rangle|^2\;, \label{C6}
\end{equation}
which has a very mild dependence on the renormalization scale. This  approximation has no solid theoretical justification, other than its simplicity. Hence, there is no reliable way of estimating corrections, which in fact appear to be rather large from comparisons between Eq. (\ref{C6}) and direct determinations from data \cite{DS1}. This poses no problem for the finite temperature analysis, where Eq.(\ref{C6}) is only used to normalize results at $T=0$, and one is usually interested in the behavior of ratios.\\
The second pillar of the QCD sum rule technique is Cauchy's theorem in the complex squared energy $s$-plane, leading to the FESR (at leading order in PQCD)
\begin{eqnarray}
&(-)^{(N-1)}& C_{2N} \langle {\mathcal{\hat{O}}}_{2N}\rangle = 4 \pi^2 \int_0^{s_0} ds\, s^{N-1} \,\frac{1}{\pi} {\mbox{Im}} \Pi_0(s)|_{\mbox{\scriptsize
{HAD}}}
\nonumber \\ [.3cm]
&-& \frac{s_0^N}{N} \left[1+{\mathcal{O}}(\alpha_s)\right] \;\; (N=1,2,\cdots) \;.\label{FESR}
\end{eqnarray}
\begin{figure}[ht]
\includegraphics[height=3.2in, width=3.7in]{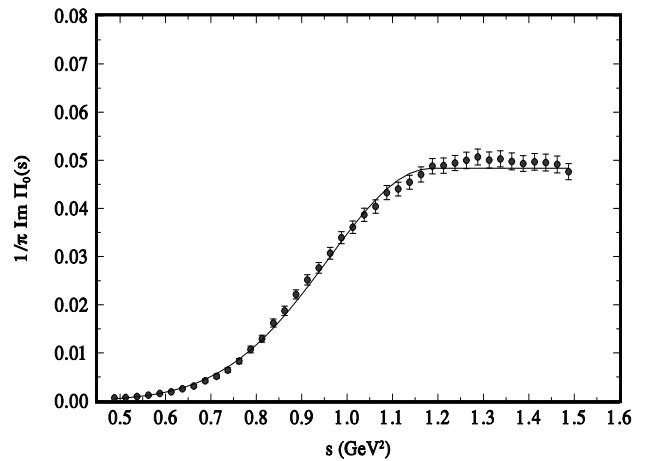}\caption{{\protect\small
{The ALEPH data in the axial-vector channel \cite{ALEPH}, and in the resonance region together with the fit, Eq.(\ref{A1}) in the region relevant to the FESR.}}}
\label{figure1}
\end{figure}
The normalization of the correlator in PQCD is
\begin{equation}
{\mbox{Im}}\, \Pi_0(s)|_{\mbox{\scriptsize
{QCD}}} = \frac{1}{4\,\pi} \left[1 + {\cal{O}}(\alpha_s(s))\right]\;. \label{NORM}
\end{equation}
At $T=0$ the radiative corrections above are known up to five-loop order, i.e. ${\cal{O}}(\alpha_s^4)$, in PQCD. Higher dimensional condensates are poorly known \cite{DS1} and thus will not be considered here.\\
\begin{figure}[hb]
\includegraphics[height=3.2in, width=3.7in]{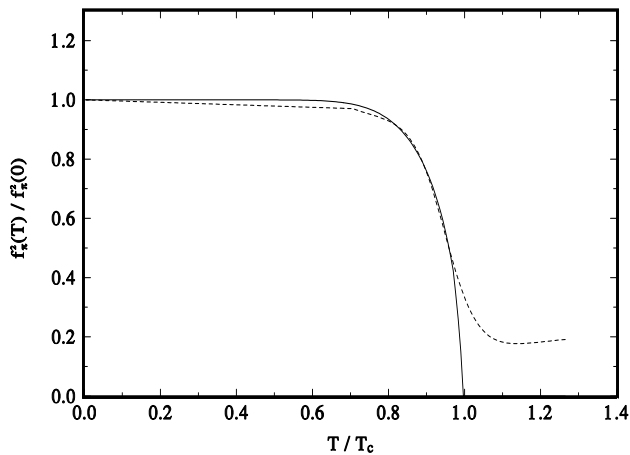}\caption{{\protect\small
{The quark condensate $\langle \bar{q} q \rangle (T)/\langle \bar{q} q \rangle(0) = f_\pi^2(T)/f_\pi^2(0)$ as a function of $T/T_c$ in the chiral limit ($m_q=M_\pi=0$) with $T_c = 197\; {\mbox{MeV}}$ \cite{QIN} (solid curve), and for finite quark masses from a fit to lattice QCD results \cite{LATTICE} (dotted curve).}}}
\label{figure2}
\end{figure}
In the hadronic sector the spectral function involves the pion pole followed by the $a_1(1260)$ resonance
\begin{equation}
{\mbox{Im}}\, \Pi_0(s)|_{\mbox{\scriptsize{HAD}}} = 2 \,\pi\, f_\pi^2 \;\delta(s) + {\mbox{Im}}\, \Pi_0(s)|_{a_1}\;, \label{HAD}
\end{equation} 
where $f_\pi = 92.21\, \pm\, 0.14 \; {\mbox{MeV}}$ \cite{PDG} is the pion decay constant, the pion mass has been neglected, and a fit to the ALEPH data \cite{ALEPH} in the resonance region gives
\begin{equation}
\frac{1}{\pi}\;{\mbox{Im}} \;\Pi_0(s)|_{a_1} = C \, f_{a_1} \; \exp{\, \left[- \left(\frac{s - M_{a_1}^2}{\Gamma_{a_1}^2}\right)\right]}\;, \label{A1}
\end{equation}
where $M_{a_1} = 1.0891\, {\mbox{GeV}}$, $\Gamma_{a_1}= 568.78\,{\mbox{MeV}}$, and $C\, f_{a_1}= 0.048326$. Using the first Weinberg sum rule as a rough estimate gives $f_{a_1} =0.073$, and thus $C =0.662$. Equation (\ref{A1}) is valid up to $s= 1.2\;{\mbox{GeV}}^2$, after which it becomes constant up to $s \simeq 1.5\;{\mbox{GeV}}^2$. This fit
together with the ALEPH data is shown in Fig.1 up to $s \simeq 1.5\, {\mbox{GeV}}^2$ (the FESR determine $s_0 = 1.44\, {\mbox{GeV}}^2$).
The pion decay constant is related to the quark condensate through the Gell-Mann-Oakes-Renner relation
\begin{equation}
2\,f_\pi^2\,M_\pi^2 = - (m_u + m_d)\langle 0| \bar{u} u + \bar{d} d|0\rangle\;. \label{GMOR}
\end{equation} 
Chiral corrections to this relation are at the 5\% level \cite{GMOR}, and at finite temperature deviations are negligible except very close to the critical temperature \cite{GMORT}.\\
The first three FESR can now be used in order to determine the PQCD threshold $s_0$, and the $d=4$ and $d=6$ condensates. These results will subsequently be used to normalize all finite temperature results. The value of $s_0$ obtained by saturating the hadronic spectral function with only the pion pole, and to leading order in PQCD, is $s_0 \simeq 0.7\, {\mbox{GeV}}^2$, as in \cite{DL1}, \cite{GATTO1}-\cite{DL3}. This value increases substantially to a more realistic $s_0 =1.15 \, {\mbox{GeV}}^2$ once the $a_1(1260)$ contribution is taken into account, and becomes $s_0 =1.44 \, {\mbox{GeV}}^2$ with PQCD to five-loop order.  Results for the leading condensates obtained from the  FESR of dimension $d=4,6$ are in good agreement with direct determinations from data \cite{DS1} at the corresponding value of $s_0$, which is to be expected as Eq.(\ref{A1}) provides a very good fit.\\
\begin{figure}[ht]
\includegraphics[height=3.2in, width=3.7in]{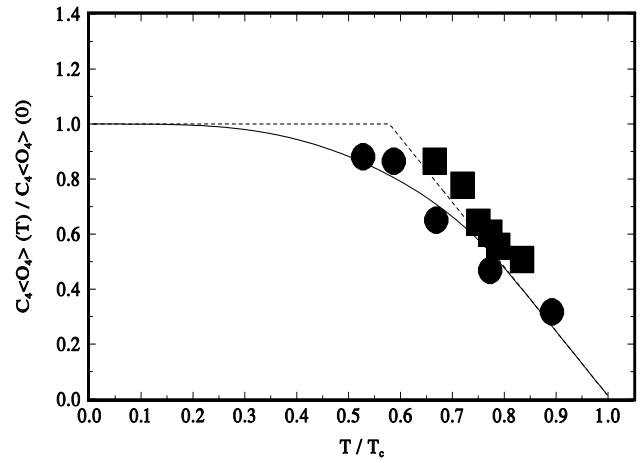}\caption{{\protect\small
{The gluon condensate $C_4\langle {\cal{O}}_4\rangle(T)/C_4\langle {\cal{O}}_4\rangle(0) $ as a function of $T/T_c$  from lattice QCD results \cite{G2_Lattice}.  Solid squares and circles correspond to 2 and 4 quark flavours, respectively, and error bars are the size of the points. The dotted curve is a fit to these data and the solid curve a smoothed fit.}}}
\label{figure3}
\end{figure}
\section{Finite Energy QCD Sum Rules at ${\bf{T\neq 0}}$}
\noindent
The extension of the QCD sum rule method to finite temperature implies a $T$-dependence in the OPE, as well as in the hadronic parameters. Regarding the former, both the Wilson coefficients as well as the vacuum condensates become temperature dependent. In particular, the strong coupling $\alpha_s(Q^2,T)$ now depends on two scales, the ordinary QCD scale $\Lambda_{QCD}$ associated with the momentum transfer, and the critical temperature scale $T_c$ associated with temperature. In principle, this poses no problems in the asymptotic freedom region and at very high temperatures, $T >> T_c$,  where PQCD can be applied. However, the QCD sum rule method approaches $T_c$ from below (starting from $T=0$), so that the presence of this second scale is problematic. No satisfactory solution to this problem exists, so that analyses must be carried out at leading order in PQCD. This circumstance is of little consequence, since basically all hadronic parameters would hardly ever be measured with the same precision as at $T=0$.
At this order in PQCD there are two thermal corrections to Eq.(\ref{NORM}), namely one in the time-like region ($q^2 >0$), the so called annihilation term which in the static limit (${\bf{q}} \rightarrow 0$) is
\begin{equation}
{\mbox{Im}} \, \Pi_0^{+}(\omega,T) = \frac{1}{4\,\pi} \left[1 - 2 \, n_F\left(\frac{\omega}{2 T}\right)\right]\;,
\end{equation}
and one in the space-like region, the so-called scattering term, originating in a cut centered at the origin on the real axis in the complex energy $\omega \equiv \sqrt{s}$-plane  of width $- |{\bf{q}}| \leq \omega \leq |{\bf{q}}|$ \cite{BS}. In the static limit  this  is given by 
\begin{eqnarray}
{\mbox{Im}} \, \Pi_0^{-}(\omega,T) &=& \frac{4}{\pi} \; \delta(\omega^2) \; \int_0^\infty \, y \; n_F\left( \frac{y}{T}\right) \; dy \nonumber \\ [.3cm]
&=&\frac{\pi}{3}\; T^2 \, \delta(\omega^2)\;, \label{ST}
\end{eqnarray}
where $n_F(z) = 1/(1 + e^{-z})$ is the Fermi thermal function, and the chiral limit was assumed.\\

\begin{figure}[hb]
\includegraphics[height=3.2in, width=3.7in]{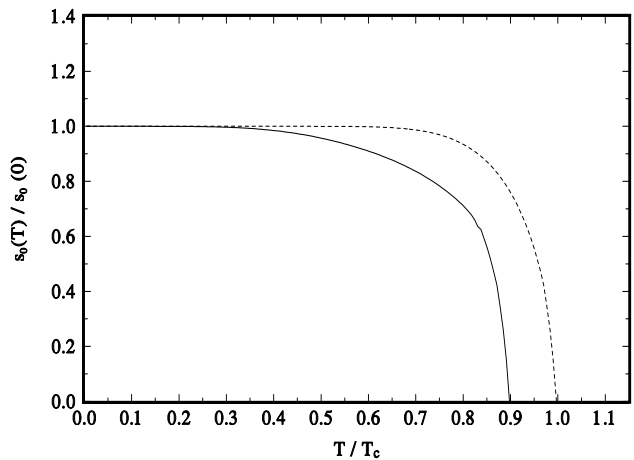}\caption{{\protect\small
{The continuum threshold $s_0(T)/s_0(0)$ signaling deconfinement (solid curve)  as a function of $T/T_c$, together with $f^2_\pi(T)/f^2_\pi(0) = \langle\bar{q} q\rangle(T)/\langle\bar{q} q\rangle(0)$ signaling chiral-symmetry restoration (dotted curve).}}}
\label{figure4}
\end{figure}
At finite temperature there are in principle additional contributions to the OPE, Eq.(\ref{OPE}), in the form of nondiagonal (Lorentz noninvariant) condensates. In the case of nongluonic operators these are highly suppressed \cite{GAMMA3},\cite{ELE}, so that they can be safely ignored. A gluonic twist-two term in the OPE was considered in \cite{Klingl}, but it is at least two orders of magnitude smaller than the standard gluon condensate at the temperatures considered here.
At dimension $d=2$ there is evidence  for the presence of a non gauge invariant condensate at high temperatures \cite{D2}. However, at the temperatures explored in the present analysis this condensate can be safely neglected.\\
In the hadronic sector and at finite temperature, masses, couplings, and widths become $T$-dependent. Hadronically stable particles, e.g. the pion, with $\Gamma(0) = 0$ develop a width , although this effect is far less pronounced than in the cases where $\Gamma(0) \neq 0$. The important parameters signaling deconfinement are the hadronic width and coupling, but not the mass. In fact, the latter is just the real part of the hadron propagator in the complex squared energy plane, while the width is its imaginary part. A vanishing mass at $T=T_c$ would not signal deconfinement, unless the width diverges at such a temperature. But then the value of the mass  becomes irrelevant, it could just as well retain its zero temperature value, or increase. This is actually what   QCD sum rule analyses show \cite{GAMMA3}, i.e. the hadronic mass is essentially constant in a wide range of temperatures, increasing or decreasing slightly very close to $T_c$, depending on the channel, and the width diverges at $T=T_c$. A notable exception are the scalar, pseudoscalar, and vector charm-anti-charm states which survive beyond $T_c$ \cite{GAMMA2}.
Finally, there is in principle a hadronic counterpart to Eq.(\ref{ST}) originating in current-pion scattering. However, in the axial-vector channel this term is loop suppressed, as the current only couples to an odd number of pions.\\
\begin{figure}[hb]
\includegraphics[height=3.2in, width=3.7in]{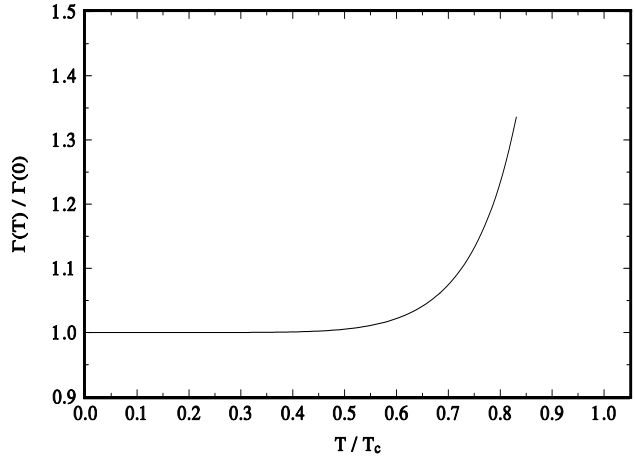}\caption{{\protect\small
{The hadronic width of the $a_1(1260)$ resonance $\Gamma_{a_1}(T)/\Gamma_{a_1}(0)$  as a function of $T/T_c$.}}}
\label{figure5}
\end{figure}
The temperature behaviour of the quark condensate, equivalently $f_\pi^2$, is shown in Fig. 2 in the chiral limit (solid curve) as determined in the framework of the Schwinger-Dyson equation \cite{QIN}, and for finite quark masses from a fit to lattice QCD results (dotted curve) \cite{DL3}, \cite{LATTICE}. The critical temperature is $T_c = 197 \; {\mbox{MeV}}$. In the sequel we concentrate on the chiral limit as we find that the FESR have solutions only up to $T \simeq 0.9 \, T_c$, a region where the quark condensate is essentially unique. The gluon condensate is shown in Fig. 3 from a fit to  lattice QCD determinations (dotted curve) \cite{G2_Lattice}, together with  a smoothed fit, both adjusted to $T_c = 197\; {\mbox{MeV}}$. This smooth fit is needed to avoid instability in the FESR due to the sharp break in the lattice QCD curve.\\
Making use of the above information the first three thermal FESR become
\begin{eqnarray}
8  \pi^2 f^2_\pi(T) &=& \frac{4}{3}  \pi^2  T^2  + \int_0^{s_0(T)}ds \,\left[1 - 2\, n_F \left(\frac{\sqrt{s}}{2 T} \right) \right] \nonumber \\ [.3cm]
&-& 4 \,\pi^2\, \int_0^{s_0(T)} ds\,  \frac{1}{\pi}\, {\mbox{Im}}\, \Pi_0(s,T)|_{a_1}
\;, \label{FESRT1}
\end{eqnarray}
\begin{eqnarray}
- C_{4}\langle {\mathcal{\hat{O}}}_{4}\rangle(T) &=& 4 \pi^2 \int_0^{s_0(T)} ds\, s \frac{1}{\pi} {\mbox{Im}}\, \Pi_0(s)|_{a_1}
\nonumber \\ [.3cm]
&-&  \int_0^{s_0(T)}ds \, s \left[1 - 2  n_F\left(\frac{\sqrt{s}}{2 T}\right)\right] ,\label{FESRT2}
\end{eqnarray}
\begin{eqnarray}
&& C_{6}\langle {\mathcal{\hat{O}}}_{6}\rangle(T) = 4 \pi^2 \int_0^{s_0(T)} ds\, s^2 \frac{1}{\pi} {\mbox{Im}}\, \Pi_0(s)|_{a_1}
\nonumber \\ [.3cm]
&-&  \int_0^{s_0(T)}ds \; s^2 \left[1 - 2  n_F\left(\frac{\sqrt{s}}{2 T}\right)\right] \;.\label{FESRT3}
\end{eqnarray}
These equations determine the continuum threshold $s_0(T)$, the coupling of the $a_1(1260)$ to the axial-vector current, $f_{a_1}(T)$, and its width $\Gamma_{a_1}(T)$, using as input the thermal quark condensate (or $f^2_\pi(T)$), the thermal $d=4,6$ condensates, and assuming the $a_1(1260)$ mass to be temperature independent, as evidenced by results in many different channels \cite{GAMMA3}, \cite{GAMMA2}, \cite{NUCLEON}.\\
\section{Results and Conclusions} 
The FESR have solutions for the three parameters, $s_0(T)$, $f_{a_1}(T)$, and $\Gamma_{a_1}(T)$, up to $T \simeq (0.85-0.90) \,T_c$, a temperature at which $s_0(T)$ reaches its minimum. An inspection of Fig.\ref{figure2} shows that at such temperatures the thermal quark condensate, or equivalently $f_\pi(T)$, is independent of whether the chiral limit (massless quarks) is assumed or not.
A short extrapolation to $T=T_c$ is to be understood for all results in the sequel.
An inspection of Eq.(\ref{FESRT1}) shows that disregarding the $a_1(1260)$ contribution, $s_0(T)$ would vanish at a lower critical temperature than $f_\pi(T)$ (or $\langle\bar{q} q\rangle(T)$). In fact, making the very rough approximation of neglecting the thermal factor $n_F(\sqrt{s}/2T)$ in the second term on the r.h.s. of Eq.(\ref{FESRT1}) leads to $s_0(T) \simeq 8 \, \pi^2\, f^2_\pi(T) - (4/3)\, \pi^2 \, T^2$. 
\begin{figure}[ht]
\includegraphics[height=3.2in, width=3.7in]{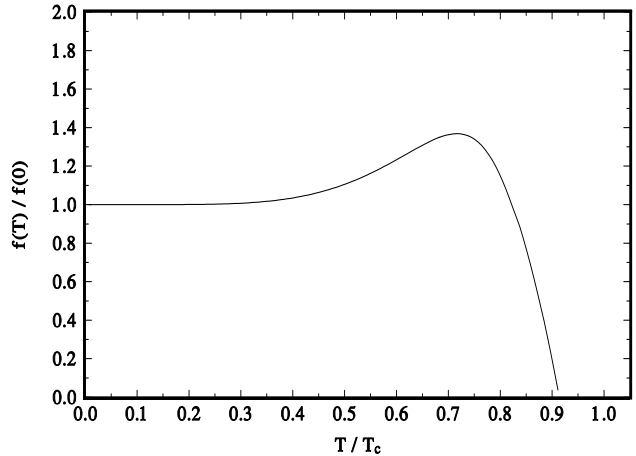}\caption{{\protect\small
{The coupling of the $a_1(1260)$ resonance $f_{a_1}(T)/f_{a_1}(0)$  as a function of $T/T_c$.}}}
\label{figure6}
\end{figure}
This feature remains valid even after including the $a_1(1260)$ in the FESR, as shown in Fig.\ref{figure4}, corresponding to the  solution for $s_0(T)$ using all three FESR. . In any case, this 10\% difference is well within the accuracy of the method. The behaviour of the width is shown in Fig.\ref{figure5}, and that of the coupling in Fig.\ref{figure6}. The rise of the width, and the fall of the coupling are indicative of a transition to a quark-gluon deconfined phase at $T = T_c$, and provide additional support for the interpretation of $s_0(T)$ as a phenomenological order parameter for quark-gluon deconfinement. It should be stressed that resonance broadening as obtained here is the result of an interplay between QCD and hadronic information. Hence, it is  directly related to quark-gluon deconfinement, in contrast to width results at finite $T$ from purely hadronic models such as e.g. the sigma model \cite{SIGMA}, which reflect a purely hadronic (absorption) effect in a medium.\\

\section{Acknowledgments} 
This work has been supported in part by NRF (South Africa), FONDECYT 1095217 (Chile), and Proyecto Anillos ACT 119 (Chile). One of us (CAD) wishes to thank Enrique Ruiz-Arriola and Eugenio Megias for correspondence on the $d=2$ condensate.

\end{document}